\documentclass[11pt]{article}
\usepackage{amssymb}
\usepackage{amsmath}
\usepackage{graphicx}
\usepackage{hep}
\usepackage{sbl_th}
\usepackage{datetime}
\usepackage{amssymb, axodraw}
\usepackage{amsmath}
\usepackage{graphicx}
\usepackage{hep}
\usepackage{color}
\newcommand{\hzero}{\ensuremath{\PHiggslightzero}} %light neutral Higgs
\newcommand{\Hzero}{\ensuremath{\PHiggsheavyzero}} %heavy neutral Higgs
\newcommand{\Azero}{\ensuremath{\PHiggspszero}} % CP-odd neutral Higgs
\newcommand{\Hpm}{\ensuremath{\PHiggspm}} %charged Higgs
\newcommand{\CP}{\ensuremath{\mathcal{C}\mathcal{P}}}

\newcommand{\bsg}{\ensuremath{\mathcal{B}(b \to s \gamma)}}
\newcommand{\pdirect}{\HepProcess{\Pphoton\Pphoton \to h}}

\begin{document}
%\pubblock

\begin{center}
\vspace{1.3cm}
\begin{center}
\vskip 1.5cm
    {\large \textsc{Single Higgs-boson production at a photon-photon
    collider:\\
    general 2HDM versus MSSM}}

    \vskip 0.7cm

\textbf{David L\'opez-Val$^{a}$ and Joan Sol\`a$^{b}$}

\vskip 0.7cm

\textit{$^{a}$ Institut f\"ur Theoretische Physik,Universit\"at Heidelberg\\
Philosophenweg 16, D-69120 Heidelberg, Germany}\\

\vskip5mm

\textit{$^{b}$ High Energy Physics Group, Dept. ECM, and Institut de Ci{\`e}ncies del Cosmos\\
Univ. de Barcelona, Av. Diagonal 647, E-08028 Barcelona, Catalonia, Spain}\\

E-mails: lopez@thphys.uni-heidelberg.de, sola@ecm.ub.es.
\vskip15mm

\end{center}

\begin{quotation}
\noindent {\large\it \underline{Abstract}}.\ \ We revisit the
production of a single Higgs boson from direct
$\Pphoton\Pphoton$-scattering at a photon collider. We compute the
total cross section $\sigma(\Pphoton\Pphoton \to h)$ (for $h=\hzero,
\Hzero, \Azero$), and the strength of the effective
$g_{\hzero\Pphoton\Pphoton}$ coupling normalized to the Standard
Model (SM), for both the general Two-Higgs-Doublet Model (2HDM) and
the Minimal Supersymmetric Standard Model (MSSM). In both cases the
predicted production rates for the $\CP$-even (odd) states render up
to $10^4$ ($10^3$) events per $500 \invfb$ of integrated luminosity,
in full consistency with all the theoretical and phenomenological
constraints. Depending on the channel the maximum rates can be
larger or smaller than the SM expectations, but in most of the
parameter space they should be well measurable. We analyze how these
departures depend on the dynamics underlying each of the models,
supersymmetric and non-supersymmetric, and highlight the possible
distinctive phenomenological signatures. We demonstrate that this
process could be extremely useful to discern non-supersymmetric
Higgs bosons from supersymmetric ones. Furthermore, in the MSSM
case, we show that $\gamma\gamma$-physics could decisively help to
overcome the serious impasse afflicting Higgs boson physics at the
infamous ``LHC wedge''.
\end{quotation}
\vskip 8mm
\end{center}

\section{Introduction}
\label{sec:intro}

Deciphering the origin of the Electroweak Symmetry Breaking (EWSB)
and the generation of masses is perhaps the most pressing unsettled
puzzle in the theory of Elementary Particles.  The Higgs
(Englert-Brout and Guralnik-Hagen-Kibble) mechanism \cite{pioners}
endows the Standard Model (SM) of Strong and Electroweak
interactions with an elegant answer, at the expense of introducing a
new (and so far unobserved) neutral, spinless, fundamental degree of
freedom. The discovery of the Higgs boson, and the study of its
phenomenological features, ranks very high in the wish list of the
experimental program currently underway at the Tevatron and the LHC
\cite{haberrev}.  Beyond its simplest description embodied by the
SM, the phenomenon of EWSB can originate from a more complicated
structure entailing a larger spectrum of Higgs bosons and a richer
pattern of interactions. The general Two-Higgs-Doublet Model
(2HDM)\,\cite{hunter,Branco11} is a trademark example of the latter,
and it is realized, in particular, by the Higgs sector of the
Minimal Supersymmetric Standard Model (MSSM) \cite{HiggsRevs}.
Should the Higgs mechanism be the option actually chosen by Nature,
it would then be mandatory to experimentally settle not only the
quantum numbers and mass spectrum of the Higgs boson(s), but also
the entire dynamics of the sector: namely, the gauge couplings of
the Higgs bosons, their Yukawa couplings to the quarks and leptons
and their own self-interactions. In this endeavor, the future
TeV-range Linear Colliders can play a key role as complementary tool
to the currently ongoing hadronic machines \cite{ILCPhysics}.

As stressed repeatedly in the literature, one particularly
interesting running mode of a linac facility is the real
$\gamma\gamma$ mode\,\cite{gammacolliders}. While the basic
operation setup for linear colliders is the head-on scattering of
high energetic electrons/positron beams, a very compelling
alternative consists in transforming such $\APelectron\Pelectron$
facility into a photon-photon (or eventually an electron-photon)
machine through Compton (back)-scattering of the original lepton
beams with laser pulses. Among the many attractive features, photon
colliders would allow to directly probe the loop-induced
$\gamma\gamma\PHiggs$ coupling, which constitutes a direct handle on
the quantum structure of the SM -- and, in fact, of any Higgs sector
extension of it, such as the general 2HDM, or the particularly
interesting case of the MSSM. However, whereas the MSSM computation
of $\sigma(\gamma\gamma\to h)$ has been dealt with on several
occasions in the literature since long ago\,\cite{2phot_mssm,previousmssm}, to
the best of our knowledge the first calculation of
$\sigma(\Pphoton\Pphoton \to h)$ in the general 2HDM is the one
presented in Ref.\,\cite{previous}, where the production of one
single Higgs boson is addressed both from the point of view of real
$\gamma\gamma$ collisions, and also within the more traditional
viewpoint of the quasi-real two-photon scattering mode at an
$\APelectron\Pelectron$ collider\footnote{See also
Ref.~\cite{doublephoton} for the study of the Higgs pairwise
production $\Pphoton\Pphoton \to hh$, and \cite{Posch10} for related
processes.}.

In this Letter, we revisit our original results \cite{previous} in
the light of the most recent and restrictive set of constraints on
the 2HDM parameter space, and we take the opportunity to closely
compare the new 2HDM results with our own independent calculation of
the corresponding $\gamma\gamma \to h$ yield for the MSSM, while
highlighting also the distinctive signatures in each case. It is
important to understand that the enhancing mechanisms in both
frameworks can be very different. While in the context of the MSSM
we expect a panoply of Yukawa, and Yukawa-like, couplings of various
kinds (including squark interactions with the Higgs bosons), whose
phenomenological implications have been exploited in the past in a
variety of important processes (see e.g.\,\cite{SUSYenhanced}), in
the case of the general 2HDM we count on alternative mechanisms.
Here we rely not only on the enhanced Yukawa couplings with Higgs
bosons, but also on the trilinear self-interactions of the latter,
whose potential effects have also been investigated in great detail
in the past, as well as recently, for different processes of Higgs
boson decay and
production\,\cite{loop1,loop2,Moretti2010,2HDMenhanced}. Worth
noticing is that these enhanced trilinear interactions are not
possible for the MSSM, a fact which may lead in principle to a
significant distinction. However, the many restrictions imposed by
perturbativity, unitarity, custodial symmetry, flavor physics,
direct searches etc. may greatly subdue the overall impact of the
enhancement sources in both frameworks, and it is not obvious how
these processes compare to each other and whether they have
realistic possibilities to be measured in the light of the present
bounds. Therefore, we believe that a fully updated comparative study
of the $\Pphoton\Pphoton \to h$ mechanism in the general 2HDM versus
the MSSM is timely and can be very useful to illustrate the
importance of the direct $\gamma\gamma$ collisions for the study of
the Higgs boson physics.

The most remarkable conclusion of this investigation is that,
despite the many theoretical and phenomenological restrictions
substantially undermining the full enhancing capabilities of the new
interactions beyond the SM, the $\Pphoton\Pphoton \to h$ processes
may definitely play a momentous role in the task of neatly
disentangling the nature of the Higgs boson(s) potentially produced
in the future TeV-class linear $\APelectron\Pelectron$ colliders
running in the $\gamma\gamma$ mode. This mode provides perhaps one
the cleanest mechanisms to study Higgs boson physics in the high
energy colliders.

\section{Phenomenological and computational setup}

The general 2HDM\,\cite{hunter} follows by extending the SM Higgs
sector with a second $SU_L(2)$ doublet with weak hypercharge $Y=+1$
and by considering the most general two-Higgs-doublet scalar field
potential that one can construct compatible with \CP-invariance and
renormalizability. Its physical spectrum contains two charged
states, $H^{\pm}$, two neutral CP-even $h^0, H^0$ (with masses
$M_{h^0}<M_{H^0}$) and one CP-odd state $A^0$. The structure of the
2HDM potential can eventually be expressed in terms of {the masses
of the physical Higgs particles ($M_{h^0}$, $M_{H^0}$, $M_{A^0}$,
$M_{H^\pm}$); the parameter $\tan \beta$ (the ratio $\langle
H_2^0\rangle/\langle H_1^0\rangle$ of the two VEV's giving masses to
the up- and down-like quarks);} the mixing angle $\alpha$ between
the two $\CP$-even states; and, finally, of one genuine Higgs boson
self-coupling, usually denoted as $\lambda_5$, which cannot be
expressed in terms of masses or other parameters of the model
\footnote{We refer the reader to Ref.\,\cite{loop1} for full details
on the model setup, notation, definitions and various constraints.}.
As for the Yukawa sector involving Higgs-quark interactions, the
absence of tree-level flavor changing neutral currents (FCNC) leads
to two main 2HDM scenarios: 1) type-I 2HDM, in which one Higgs
doublet couples to all quarks, whereas the other doublet does not
couple to them at all; 2) type-II 2HDM, {where one doublet couples
only to down-like quarks and the other doublet just to up-like
quarks}. Other flavor structures are also conceivable and have
indeed attracted a growing attention in the recent years
\cite{newflavor}, but we will stick here to just the two
aforementioned leading 2HDM models, which traditionally represent
the two canonical options.

The very same {Higgs spectrum} emerges naturally from the MSSM,
although SUSY constraints narrow the free parameters of the Higgs
sector down to 2, usually taken to be $\tan\beta$ and $M_{\Azero}$.
The corresponding Yukawa sector mimicks that of a type-II one,
although of a very restricted sort -- enforced again by SUSY
invariance\,\cite{hunter}. Most significantly, while the generic
2HDM allows triple (3H) and quartic (4H) Higgs self-interactions to
be largely enhanced, in the MSSM these Higgs self-couplings are
restrained to be purely gauge-like. The phenomenology of such
potentially large 3H self-interactions has been actively
investigated at $\APelectron\Pelectron$ linear colliders within a
manifold of processes, and compared to their counterpart processes
in the MSSM. These analyses include e.g. the tree-level production
of triple Higgs-boson final states \cite{giancarlo}; the double
Higgs-strahlung channels $hh\PZ^0$ \cite{arhrib}; and the inclusive
Higgs-pair production via gauge-boson fusion \cite{neil}. Also their
fingerprint at the quantum level, in the form of large quantum
effects, has been comprehensively reported in \cite{loop1,loop2}.
All the abovementioned dynamical features also play a sensible role
in the structure of the $\Pphoton\Pphoton\PHiggs$ coupling, as we
shall see hereafter, and could not only entail hints of non-standard
Higgs boson physics, but also a handle on the SUSY versus non-SUSY
nature of a possible extended Higgs sector.

Our study  of the process $\pdirect$ is accomplished in
correspondence with the most stringent experimental and theoretical
constraints that restrict the allowed regions in the 2HDM and the
MSSM parameter spaces. They stem fundamentally from the requirements
of perturbativity, unitarity and vacuum stability, as well as from
the EW precision data, the low-energy flavor-physics inputs and the
Higgs mass regions ruled out by the LEP and Tevatron direct
searches. Several studies in the literature provide a dedicated
account on these topics ~\cite{constraints_general,superiso}. A more
detailed description of the role of these constraints in the context
of our analysis may be found e.g. in Ref.~\cite{loop1}. Let us
stress, in particular, the critical role of perturbative unitarity,
which enforces a limit on the strength of the {Higgs
self-interactions}. In the present study we employ the most
restrictive set of conditions proposed in Ref.\,\cite{unitarity} and
discuss their impact with respect to the simplified approach that
was first employed in our preliminary study of Ref.~\cite{previous}.
Tight bounds also follow from the radiative $B$-meson decay ($b \to
s\gamma$), as well as from the $B_d^0 - \bar{B}_d^0$ mixing (which
was not considered in ~\cite{previous}). While the former basically
defines a lower bound on the charged Higgs mass $M_{H^{\pm}} \gtrsim
300$ GeV for $\tan \beta \ge 1$ \cite{constraints_general} (which
only applies to type-II 2HDM, but not to type-I), the latter
strongly disfavors the regions of $\tan\beta \lesssim 1$, for both
type-I and type-II 2HDM and, in general, tends to enforce $\tan\beta
\gtrsim 2$ for light charged Higgs bosons (viz. $M_{\PHiggs^{\pm}}
\sim 100-150 \,\GeV$)\,\cite{superiso}. In our actual calculation we
have included a fairly exhaustive collection of constraints by
combining the packages \textit{2HDMCalc} \cite{2hdmcalc},
\textit{SuperISO} \cite{superiso} and \textit{HiggsBounds}
\cite{higgsbounds}, altogether with several complementary in-house
routines. As for the algebraic calculation of the $\pdirect$
cross-section,  we have made use of the standard computational
software \textit{FeynArts}, \textit{FormCalc} and \textit{LoopTools}
\cite{feynarts}. The Photon Luminosity functions, which account for
the effective $e^{\pm}\to\gamma$ ``conversion'' of the primary linac
beam, are taken from the package \texttt{CompAZ} \cite{compaz}.

\begin{table}[t!]
\begin{center}
\begin{tabular}{|c||c|c|c|c|}
\hline
        2HDM & $M_{\hzero}$ (GeV)& $M_{\Hzero}$ (GeV)& $M_{\Azero}$ (GeV)& $M_{\Hpm}$(GeV)  \\
\hline\hline
Set I & $115$ &  $165$ &  $100$  & $105$  \\
Set II & $200$ &  $250$ &  $290$  & $300$  \\
\hline
\end{tabular}
\end{center}
\caption{\footnotesize{Sets of 2HDM Higgs boson masses used
throughout the calculation. Owing to the $\mathcal{B}(b \to
s\gamma)$ constraints on
$M_{\Hpm}$\,\,\cite{Misiak:2006zs}, Set I is only possible
for type-I 2HDM's, whereas Set II is possible for both type-I and
type-II. The mass sets are enforced to satisfy the custodial
symmetry bound $|\delta\rho|<10^{-3}$ -- cf. Ref.\cite{loop1}.}}
\label{tab:mass2hdm}
\end{table}

\section{Numerical analysis}

We shall next provide the main numerical results. For lack of space,
in this Letter we cannot furnish analytical expressions for the
calculation of the corresponding cross-sections. For explicit
details, in particular for the complete set of Feynman diagrams and
for the formulae that relate the basic ``partonic''
$\sigma(\Pphoton\Pphoton\to h)$ cross-section to the the total
averaged $\gamma\gamma$ cross-section
$\langle\sigma_{\gamma\gamma\to h}\rangle(s)$ (unpolarized and
convoluted with the differential luminosity distribution) as a
function of the linac center of mass energy $\sqrt{s}$, we refer
again the reader to our previous study of Ref.~\cite{previous}.
Furthermore, a detailed exposition of all the relevant pieces of the
2HDM interaction Lagrangian is given e.g. in our notation
in\,\cite{loop1}. The MSSM interactions are summarized e.g. in
\cite{hunter}.

\subsection{$\gamma\gamma \to h$ within the 2HDM}

Let us begin by revisiting the behavior of the total averaged
cross-section $\langle\sigma_{\gamma\gamma\to h}\rangle(s)$, as well
as of the relative strength of the effective $\Pphoton\Pphoton h$
interaction normalized to the SM, $r \equiv g_{\Pphoton\Pphoton
h}/g_{\Pphoton\Pphoton H}^{\rm SM}$, in the framework of the 2HDM.
In this context, the $\Pphoton\Pphoton h$ effective vertex is
generated at the quantum level through a rather complicated
numerical interplay of the contributions from fermion,
$\PW^{\pm}$-boson and charged Higgs boson loops, which include the
trilinear self-interactions  $\hzero\PHiggs^+\PHiggs^-$ and
$\Hzero\PHiggs^+\PHiggs^-$ -- see Fig.\, 2 of Ref.\,\cite{previous}.
In the MSSM case, we additionally have the squark and slepton loop
contributions. Already from the dynamics of {the $\Pphoton\Pphoton
h$ coupling} in the SM, we know that the contribution of the
(transverse components of the) gauge bosons are large and of
opposite sign to the fermion and the {scalar (namely the Goldstone
boson)} loops \cite{ellis}. The very same interference pattern
occurs in the 2HDM, and causes the phenomenological features to be
critically sensitive to the charged Higgs boson couplings.

\begin{figure}[t!]
\begin{center}
\begin{tabular}{c}
\vspace{0.3cm}
\includegraphics[scale=0.55]{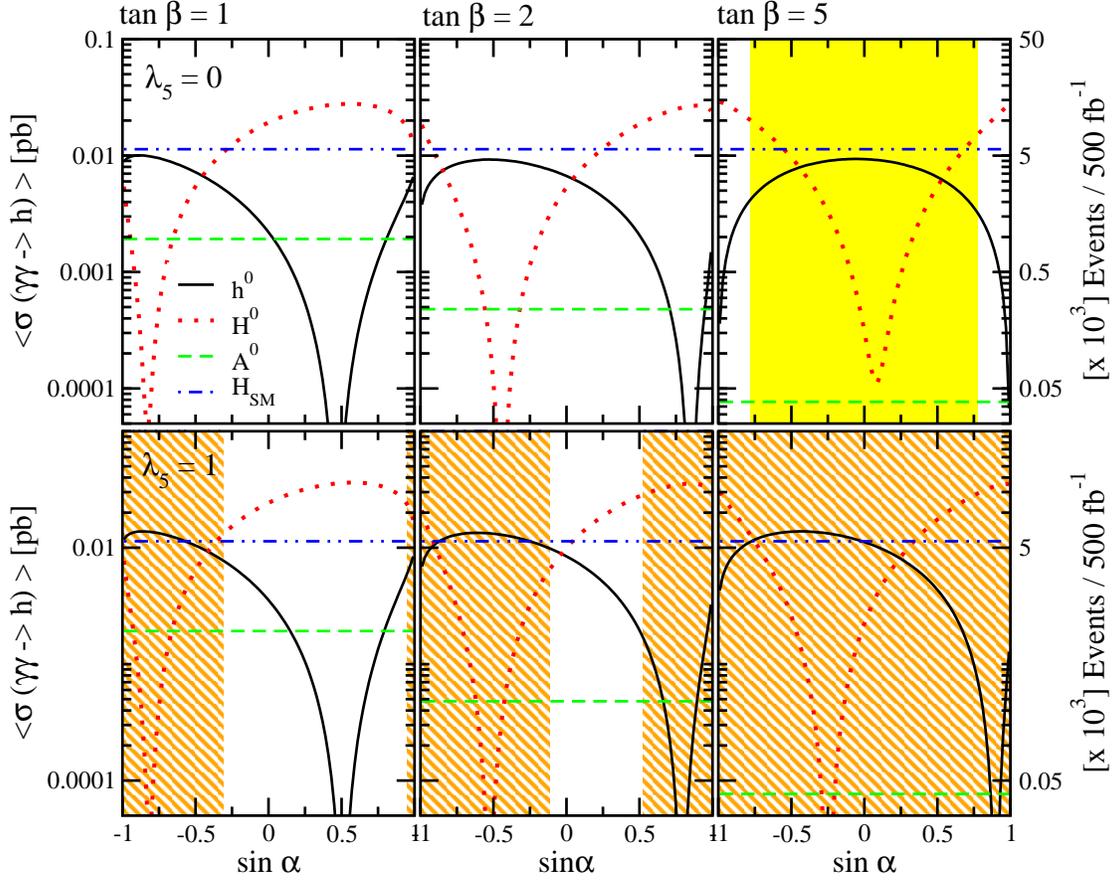}
 \end{tabular}
\caption{\footnotesize{Total averaged cross-section
$\langle\sigma_{\gamma\gamma\to \hzero}\rangle(s)$ for
$\sqrt{s}=500$ GeV, and number of Higgs boson events, as a function
of $\sin\alpha$. (Although the \CP-odd production channel
$\gamma\gamma\to\Azero$ does not depend on this parameter, it is
included for completeness.) Shown are the resulting cross-sections
for the SM (dash-dotted horizontal line at $\sigma_{\rm SM}\simeq
0.011$ pb for $M_{H_{\rm SM}}=115$ GeV), and the corresponding 2HDM
ones for Higgs boson masses as in Set I, for $\lambda_5=0$ (top
panels) and $\lambda_5 = 1$ (bottom panels), and for three values of
$\tan\beta$. Notice that the characteristic suppression of the Higgs
production rate (which takes place at different regions in the
parameter $\sin\alpha$ for each \CP-even channel) is a signature of
the complementarity of the $\hzero\PHiggs^+\PHiggs^-$ and
$\Hzero\PHiggs^+\PHiggs^-$ self-couplings (cf. Table II of
Ref.\cite{loop1}). The shaded (yellow) area in the $\tan\beta=5$
case is excluded by unitarity, {while the dashed (orange) regions in
the bottom panels are disallowed by the vacuum stability
conditions}. {Let us also underline that the $\tan\beta = 1$ case is
included to better assess the dependence of the cross section as a
function of this variable, although the constraints stemming from
$B_d^0 - \bar{B}_d^0$ exclude it (see the text and the left panel of
Fig.\,\ref{fig:compare1_2hdm}).}}}\label{fig:oversa2hdm}
\end{center}
\end{figure}

\begin{figure}[t!]
\begin{center}
\includegraphics[scale=0.55]{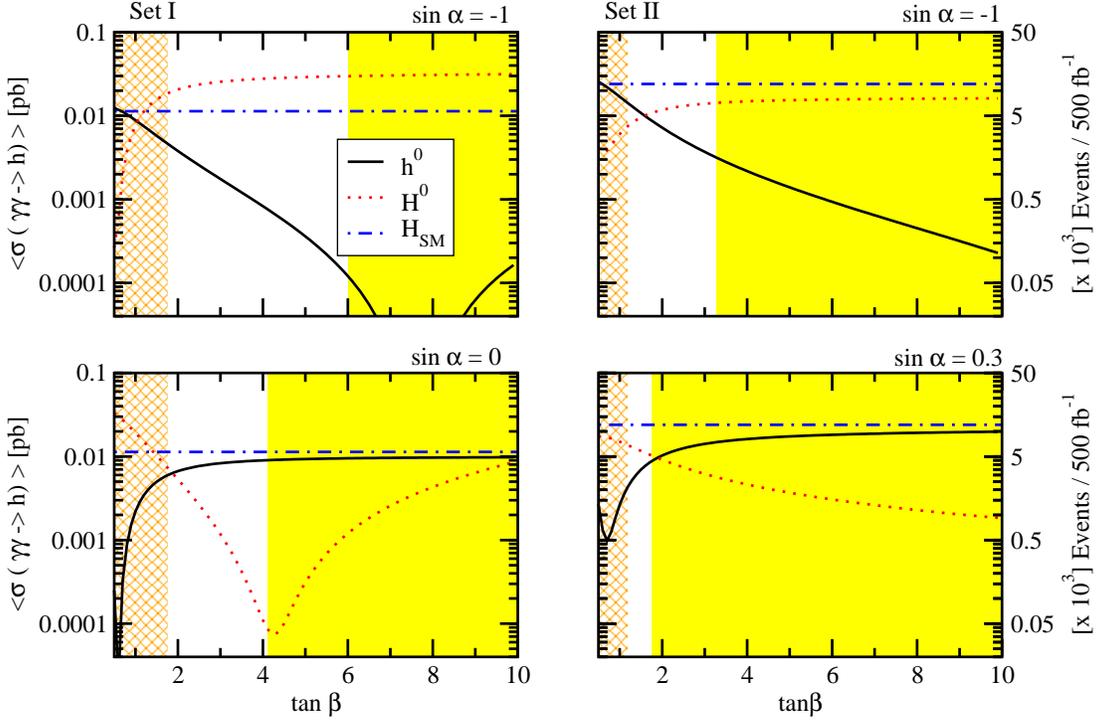}
\caption{\footnotesize{Cross-section $\langle\sigma_{\gamma\gamma\to
\hzero}\rangle(s)$ ($\sqrt{s}=500$ GeV) and number of Higgs boson
events, as a function of $\tan\beta$. We plot the \CP-even channels
only, and compare the resulting cross-sections for the SM and the
2HDM by fixing the remaining Higgs boson masses as in Sets I and II,
for $\lambda_5=0$ and different choices of $\sin\alpha$ (as
indicated in the figure). The excluded $\tan\beta$ range due to
unitarity (yellow shaded area) and $B_d^0 - \bar{B}_d^0$ mixing
(orange crossed area) are explicitly
indicated.}}\label{fig:overtb2hdm}
\end{center}
\end{figure}

In Figures~\ref{fig:oversa2hdm}-\ref{fig:overtb2hdm} we display the
evolution of the total averaged single Higgs boson cross-section
$\langle\sigma_{\gamma\gamma\to h}\rangle(s)$ at a linac center of
mass energy of $\sqrt{s}=500$ GeV, as a function of $\sin\alpha$ and
$\tan\beta$ respectively. Notice that {while in
Fig.~\ref{fig:oversa2hdm} we dwell} on Set I of Higgs boson masses
(cf. Table~\ref{tab:mass2hdm}) and compare the cases $\lambda_5 = 0$
and $\lambda_5 = 1$, in {Fig.~\ref{fig:overtb2hdm} we use} both
Higgs boson mass sets (I and II in Table~\ref{tab:mass2hdm}), but
concentrate on the setting $\lambda_5 = 0$ only. In focusing on the
latter case, we place ourselves in a scenario in which the overall
size of the relevant 3H self-interactions is modulated solely by the
actual $\sin\alpha$ and $\tan\beta$ values, along with the Higgs
boson masses\,\footnote{For a detailed list of trilinear self-Higgs
boson vertices in the general 2HDM, see e.g. Table II of
Ref.\cite{loop1}.}. Our moderate choices of $\lambda_5$ are
motivated by the most restrictive set of unitarity constraints that
we are using here~\cite{unitarity}. These constraints no longer
allow $\lambda_5$ values as large as $|\lambda_5|\gtrsim 10$, for
which the trilinear effects are very conspicuous, in fact the
leading ones\,\cite{previous}. Here we will not consider
the large $\lambda_5$ scenario anymore, and we shall instead focus
on mild values of order $|\lambda_5|={\cal O}(1)$, which fall well
within the regions permitted by unitarity. An example is the case
$\lambda_5=1$ studied in Fig.\,\ref{fig:oversa2hdm}. It is
worthwhile stressing that, for moderate values $|\lambda_5| \gtrsim
1$, we meet in general a peculiar situation whereby the contribution
from the trilinear coupling attains just the critical size which is
able to partly counterbalance the rest of the quantum effects (viz.
the loop effects triggered by the gauge bosons and the fermions with
enhanced Yukawa couplings); and as a result we encounter a
destructive interference scenario in most of the parameter space of
the 2HDM. {Let us recall, too, that the $\lambda_5 > 0$ regions tend
to be disfavored by the vacuum stability conditions -- which become
even more restrictive with growing values of $\tan\beta$ (cf. the
excluded areas in the lower panels of Fig.~\ref{fig:oversa2hdm}). }
% \newtext{It is for these reasons} that we shall
% set $\lambda_5=0$ throughout most of the present analysis.
Remarkably enough, even within this more restricted context we find
very significant potential sources of new Higgs boson physics. In
particular, the size of the cross-sections stays well within the
measurable range {and exhibits trademark phenomenological features,
as we shall analyze in what follows}.

Within this setup, Figs.~\ref{fig:oversa2hdm}-\ref{fig:overtb2hdm}
illustrate the results for the light ($\hzero$) and the heavy
($\Hzero$) neutral CP-even Higgs bosons, including also the \CP-odd
state ($\Azero$), and compare the obtained rates from the 2HDM with
the SM prediction for $M_{\PHiggs_{SM}} = M_{\hzero}$. From these
plots we can easily read off the following relevant facts: i) the
maximum cross sections may render $\sigma=\mathcal{O}(10^{-2})\,
\picobarn$; ii) the optimal $\hzero$ and $\Hzero$ event rates are
largely complementary to each other, as a result of the inverse
correlation of the respective self-interactions
$\hzero\PHiggs^+\PHiggs^-$ and $\Hzero\PHiggs^+\PHiggs^-$ (once more
we refer to Table II of Ref.\cite{loop1}), which trigger the large
suppressions (``dips'' confronted with ``cusps'') visible in the
plots (e.g. quite notably in Fig.\,\ref{fig:oversa2hdm}). We recall
that their origin can be traced back to the destructive interference
operating between the fermion, gauge boson and charged
Higgs-mediated one-loop contributions to
$g_{\hzero\Pphoton\Pphoton}$; and iii) the maximum cross-section for
the \CP-odd state $\Azero$ is significantly smaller than that of the
$\CP$-even states (at least ten times smaller) but it does not get
suppressed with $\sin\alpha$. For example, in the  $\tan\beta=1$
case indicated in Fig.\,\ref{fig:oversa2hdm} it may lead to $\sim
10^3$ events per 500 $\invfb$ of integrated luminosity. For larger
values of $\tan\beta$, however, the event rate decreases to the
$\sim10^2$ level or below.

It is encouraging to see that, for the \CP-even states, the
cross-sections can be quite sizeable away the suppressing dips in
Figures~\ref{fig:oversa2hdm}-\ref{fig:overtb2hdm}, where they can
render a few thousand events for $h^0$, and up to ten thousand
events for $H^0$, per 500 $\invfb$ of integrated luminosity.
Admittedly in some cases the combination of unitarity and $B_d^0 -
\bar{B}_d^0$ mixing constraints {enforces a relatively} narrow
region for the allowed parameter space, but in general it is still
sufficiently large. Also remarkable is the fact that while the
obtained rates for $\Pphoton\Pphoton \to \hzero$ tend to lie
slightly below their SM counterparts, the $\Pphoton\Pphoton \to
\Hzero$ channel can have instead a cross-section larger than the SM
case. This is a reflect of the behavior $\sigma(\Pphoton\Pphoton \to
h) \sim M^4_h/M_W^2$ in the general 2HDM, which implies that
$\sigma(\Hzero) > \sigma(\PHiggs_{SM})$ since $M_{\Hzero}
> M_{\PHiggs_{SM}}\equiv M_{\hzero}$.

\begin{figure}[t]
\begin{center}
\begin{tabular}{cc}
\includegraphics[scale=0.90]{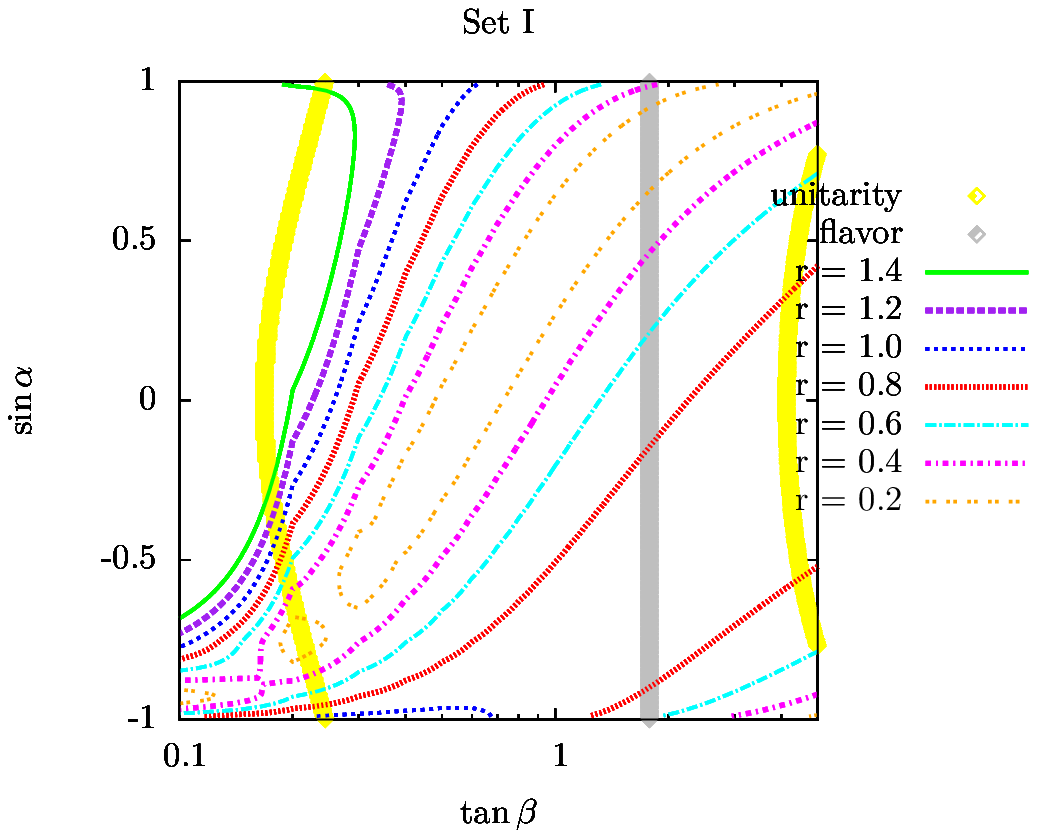} &  \includegraphics[scale=0.90]{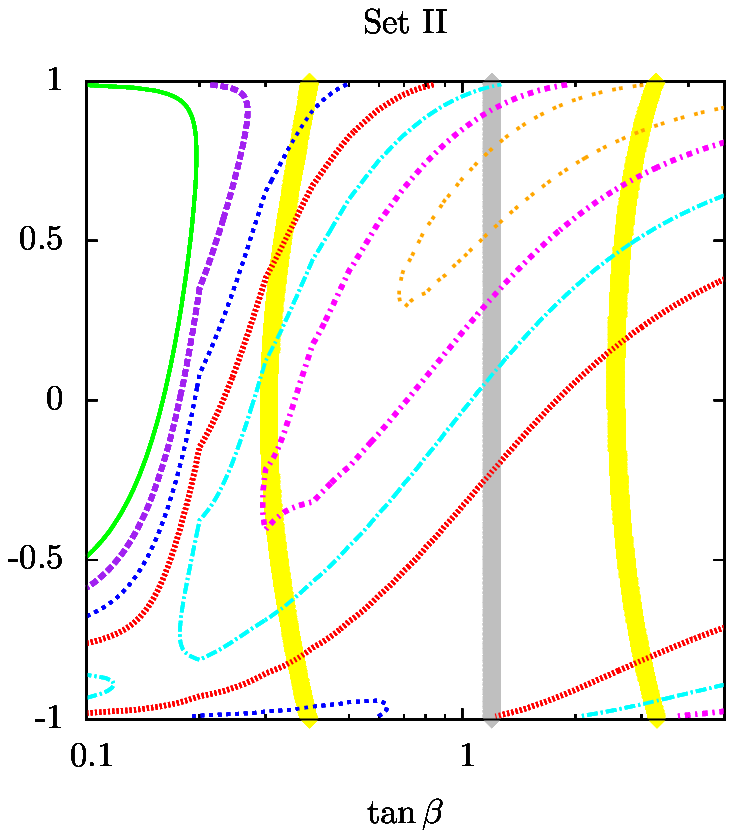}
\\ & \\
(a) & (b) \end{tabular} \caption{\footnotesize{The ratio $r \equiv
g_{\Pphoton\Pphoton \hzero}/g_{\Pphoton\Pphoton H}^{\rm SM}$
measuring the effective $\Pphoton\Pphoton\hzero$ coupling strength
in the 2HDM as compared to the SM, as a function of $\sin\alpha$ and
$\tan\beta$, for Sets I and II of Higgs boson masses in
Table~\ref{tab:mass2hdm}. The results have been obtained {by setting
$\lambda_5 = 0$.} The yellow bands depict the lower and upper bounds
on $\tan\beta$, out of which the restrictions of perturbative
unitarity are violated. In turn, the grey band stands for the lower
bound (at $3\sigma$ C.L.) enforced by $B^0_d-\bar{B}^0_d$ mixing.
The allowed region in the plots therefore is the one lying between
the grey band and the rightmost yellow band.}}
\label{fig:compare1_2hdm}
\end{center}
\end{figure}

{How does the relative size of the 2HDM cross-sections versus the SM
ones compare to the value of the ratio of the effective couplings
$\gamma\gamma h$ in both models, i.e. $r=g_{\Pphoton\Pphoton
h}/g_{\Pphoton\Pphoton H}^{\rm SM}$?} In Ref.~\cite{previous} it was
pointed out that, in the case of a type-I 2HDM, an enhancing effect
up to $r \simeq 4$ could be reached for relatively light charged
Higgs bosons (as e.g. in Set I) and large enough 3H
self-interactions -- the optimal region being $\lambda_5 ~\sim -20 $
and $\tan\beta ~\sim 1$. As we have repeatedly emphasized, in the
present analysis we adopt a more conservative perspective and hence
stick to a specific, and more restrictive, set of unitarity
constraints \cite{unitarity}. Their net effect is to pull down the
maximum strength of the $h\PHiggs^+\PHiggs^- (h=\hzero,\Hzero)$
self-coupling by a factor of roughly 4, meaning that the new
maximally allowed values of the relative coupling strength are
$r\gtrsim 1$. Figure~\ref{fig:compare1_2hdm} displays a detailed
view on how $r$ evolves in the $(\tan\beta, \sin\alpha)$ plane,
again under the assumption that $\lambda_5 = 0$ and for the same
Higgs boson mass sets. It is instructive to compare that figure with
Fig. 5 (and Table 2) of Ref.\,\cite{previous}, where we explored the
influence of $\lambda_5$ within a more relaxed set of unitarity
conditions. The reduction by a factor of $\sim 3-4$ becomes evident.

At first sight, one would expect that such reduction should
translate into a depletion of the maximum cross sections by a factor
roughly of $r^2\sim 10-20$. In practice, however, the suppression
turns out to be larger as a consequence of the aforementioned
interference between the charged Higgs boson, fermion and gauge
boson-mediated one-loop diagrams. Consequently, the potentially
distinctive imprint of type-I 2HDM, in the form of a boost (up to a
factor 10) with respect to the SM predictions fades away if we apply
the more restrictive set of unitarity conditions, as we do in the
present study. Fortunately, other distinctive phenomenological
signatures may come into play.

%%%%%%%%%%%%%%%%%%%%%%%%%%%%%%%%%%%%%%%%%%%%%%%%%%%%%%%%%%%%%%%%%%%%%%%%
\begin{figure}[t!]
\begin{center}
\includegraphics[scale=0.5]{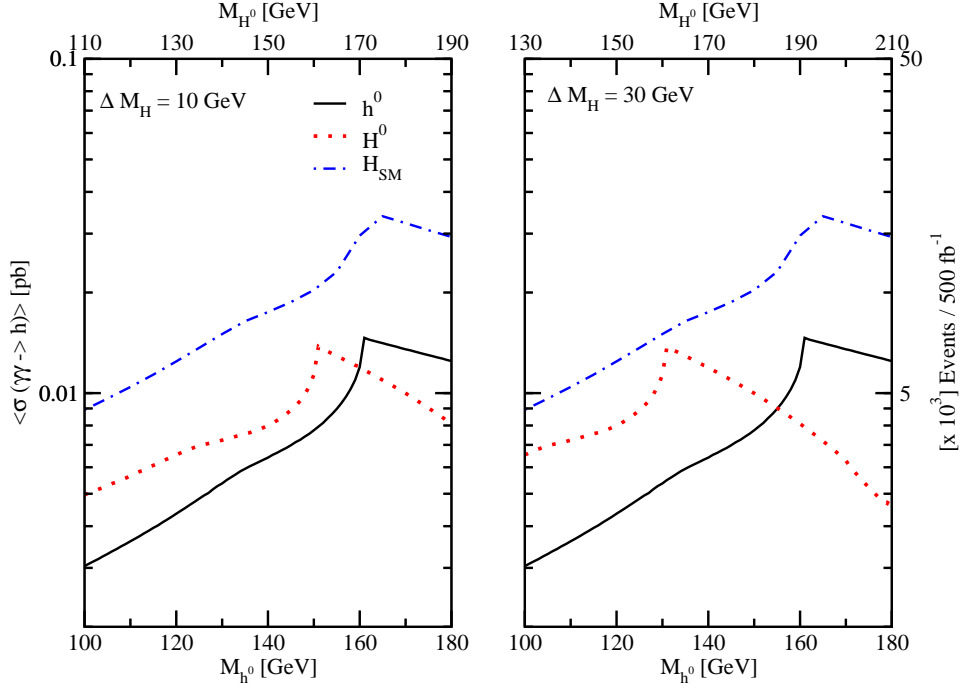}
\caption{\footnotesize{Total cross-section
$\langle\sigma_{\gamma\gamma\to h}\rangle(s)$ for $\sqrt{s}=500$ GeV
and number of Higgs boson events, as a function of the CP-even Higgs
boson masses ($M_{\hzero}$, in the lower X-axis, and $M_{\Hzero}$,
in the upper X-axis). The mass splitting between the two states is
kept at $10\,\GeV$ (left panel) and  $30\,\GeV$ (right panel). The
Higgs boson masses are as in Set I, and for $\sin\alpha = 0.30$,
$\tan\beta=2$ and $\lambda_5=0$. The SM cross-section is also
included (dash-dotted blue line). Remarkably, the two \CP-even
states (solid and dotted lines) could be simultaneously accessible
in the general 2HDM. }} \label{fig:1}
\end{center}
\end{figure}

Indeed, the relevant phenomenological signs may reside in the
parameter region in which the combination of non-standard
gauge/Yukawa couplings of the 2HDM stamp a fingerprint on the Higgs
boson production cross section, therefore far from the regions where
the triple Higgs self-interactions alone dominate the loop-induced
$\gamma\gamma h$ coupling.
%and induce their relative suppression with respect to the SM.
As we have seen, this implies low values of $\lambda_5$ and
$\tan\beta$ for a certain range of $\sin\alpha$. Notice, first of
all, the existence of rather wide regions of the parameter space for
which the $\pdirect$ cross-section departs from its SM counterpart
($\sigma_{\rm SM} \simeq 11\,\femtobarn$ for a SM Higgs mass of
$M_{H_{\rm SM}} = 115\,\GeV$, as in Set I). {These regions are
characterized by a sizable reduction -- at the level of $-10\%$ to
$-60\%$ -- of the loop-induced $\Pphoton\Pphoton\PHiggslightzero$
interaction in most of the $\sin\alpha - \tan\beta$ plane},  again
due to the destructive {interference modulated by the Higgs boson
self-coupling $\hzero\PHiggs^+\PHiggs^-$.
% Actually, in the general 2HDM large
% suppressions of the ratio $r$ can be expected, say at the level
% $r\lesssim 0.2$ (cf. Fig.\, \ref{fig:compare1_2hdm}).
On the other hand,} augmented contributions with respect to the SM
value (i.e. $r>1$) are only possible, at least theoretically, within
a very constrained range: $\tan\beta \sim 0.2-0.3$  (already
bordering the unitarity and perturbativity limit). Here $r$ can
reach $\sim 1.1 - 1.4$ (entailing {cross-sections up to $20\%$
larger than the SM ones}) driven by the Higgs-top Yukawa coupling,
which evolves as $\sim 1/\tan\beta$ and therefore becomes enhanced
in that range. Unfortunately, this region of parameter space is
essentially ruled out by the experimental constraints dictated by
$B_d^0 - \bar{B}_d^0$ mixing\,\cite{superiso}, which hold for all
possible Higgs-fermion coupling patterns. (Actually, the $3\sigma$
exclusion region extents up to values of $\tan\beta\sim 2$ for light
charged Higgs boson masses, as shown in
Fig.~\ref{fig:compare1_2hdm}a for Set I). A very similar picture is
encountered for Set II (see Fig.~\ref{fig:compare1_2hdm}b), although
the unitarity constraints become {now more stringent, due to the
presence of heavier Higgs bosons. As a consequence, the allowed
regions for which the effective $\gamma\gamma\Hzero$ departs
significantly from $r = 1$ cover a smaller patch of the $\tan\beta -
\lambda_5$ parameter space.} However, for Set II the lower bound on
$\tan\beta$ dictated by $B_d^0 - \bar{B}_d^0$ mixing is smaller:
$\tan\beta\gtrsim 1$ (cf. Fig.~\ref{fig:compare1_2hdm}b). Let us
also point out that type-I and type-II models are essentially
indistinguishable from this point of view. This is an indication
that both the Higgs-top quark coupling and the Higgs couplings to
the gauge bosons, which are the relevant interactions in this
domain, are common for both types of models.

The upshot of {our analysis so far} is that the task of spotting a
``tail of subleading effects'' triggered by the non-SM
``Yukawa-gauge'' sector of the theory should be perfectly feasible.
Even if it might not enable discerning the particular type of 2HDM,
the missing number of events could be a vigorous hint of a smoking
gun -- namely, of Higgs boson physics beyond the SM. This is of
course under the assumption that the overall Higgs production {rates
lie} only moderately below the SM predictions. Should the depletion
be much larger, the actual missing number of events might not be
enough to disentangle the signal from the dominant background
process $\gamma\gamma \to b\bar{b}$.

Finally, in Fig.\,\ref{fig:1} we illustrate a very interesting
phenomenological situation that {could be particularly
representative} of genuine 2HDM physics. We consider the
simultaneous production of two \CP-even Higgs bosons with moderate
mass splittings of $\Delta M_H=10$ GeV and  $\Delta M_H=30$ GeV. We
focus our study around a mass region that comprises the upper mass
bound that applies on the lightest \CP-even Higgs boson $\hzero$ in
the MSSM, i.e. $M_{\hzero}^{\rm max}\simeq 115-140$ GeV. {The
results show that it is perfectly possible to produce simultaneously
the two \CP-even Higgs states with similar masses in the general
2HDM, and both with large event rates of order $\sim 10^3$ for the
usual integrated luminosity of 500 $\invfb$ -- and {for relatively
light (as in Set I) or heavy (as in Set II) Higgs boson spectra
alike}.} This situation is impossible to realize in the MSSM, and
therefore it would be a very distinctive signature of
non-supersymmetric Higgs boson physics in a photon collider. In the
next subsection, we dwell on the MSSM case in more detail.

\subsection{$\gamma\gamma \to h$ within the MSSM}

In a similar vein, we briefly address now the single $\gamma\gamma$
production of Higgs bosons in the MSSM. While the general 2HDM case
was first studied only very recently\,\cite{previous}, the MSSM
process has received a lot more of
attention\,\cite{2phot_mssm,previousmssm}. Here we revisit the
supersymmetric process in order to better compare with our detailed
account of the general 2HDM case. The bottom-line of the MSSM
studies on this process can be summarized as follows: in the most
favorable situations, the relative effective strength of the
$\gamma\gamma h$ vertex with respect to the SM can reach up to
$r\simeq\sqrt{2}\simeq 1.4$. There are basically two conditions
under which this enhancements could be implemented: i) a large mass
splitting between the chiral components of the squarks, in
particular the stops -- one of them being as light as possible; and
ii) a large Higgs-squark Yukawa-like coupling, which means, for the
stop in particular, a large value of the trilinear coupling $A_t$.
{The foresaid mass splitting can essentially be traced back to the
soft-SUSY breaking pattern in the squark mass sector which,
following standard conventions, can be written in terms of the mass
matrix}

\begin{equation}
M^2_{\tilde{Q}}\, = \left(\begin{array}{cc}
M^2_{\tilde{Q}L} + m^2_f + \cos 2\beta\,(T_3^{f_L} - Q_f\,\sin^2\theta_w)\,M_Z^2 & m_f\,M_{LR}^f \\
m_f\,M_{LR}^f & M^2_{\tilde{Q} R} + m_f^2 + \cos 2\beta Q_f \sin^2\theta_w M_Z^2
\end{array}
\right)
\label{eq:sfermionmass2},
\end{equation}
{where $M_{\tilde{Q}_{L,R}}$ denote the soft-SUSY breaking masses
for the left-handed (resp. right-handed) squark fields; while the
off-diagonal pieces correspond to $M^u_{LR} = A_u - \mu\cot\beta$
and $M^d_{LR} = A_d - \mu\tan\beta$. If the mass splitting $\Delta
m_{\tilde{f}}=m_{\tilde{f}_1}-m_{\tilde{f}_2}$ between the two mass
eigenvalues is significant, this generates an asymmetry in the loop
contributions to $\gamma\gamma\to h$ induced by each of the squark
components and allows a {neat overall yield} with a strength
comparable to the gauge boson and the fermion-mediated
counterparts.}

%\begin{figure}[htb]
\begin{figure}[t]
\begin{center}
\includegraphics[scale=0.6]{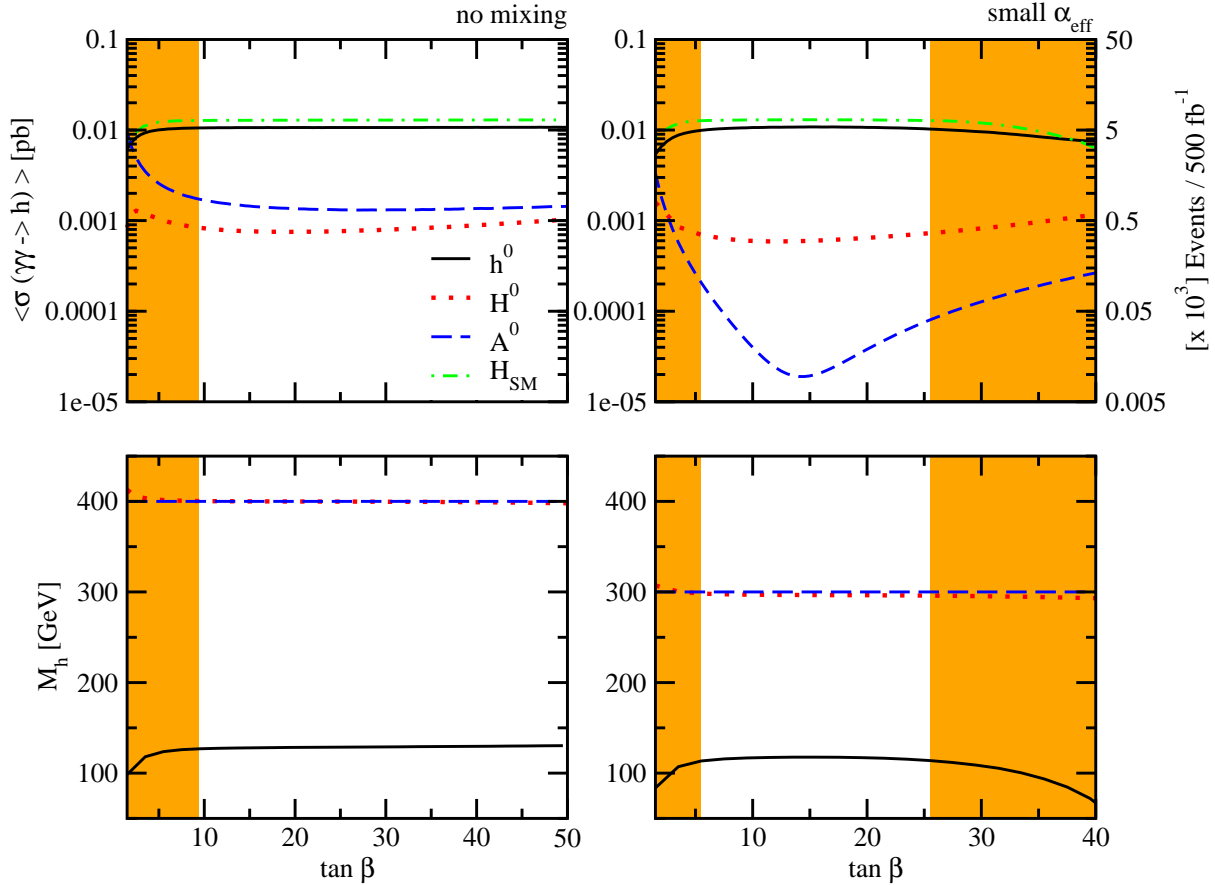}
\caption{\footnotesize{Cross-section $\langle\sigma_{\gamma\gamma\to
\hzero}\rangle(s)$ and number of Higgs boson events as a function of
$\tan\beta$. We plot the resulting cross-sections for the MSSM
within the benchmark scenarios quoted in Table~\ref{tab:mssmparam},
and compare them to the SM. In the bottom panels we account for the
the light (versus heavy) neutral \CP-even MSSM Higgs boson masses as
a function of $\tan\beta$. The shaded bands stand for the excluded
mass regimes. The center of mass energy is fixed at {$\sqrt{s} =
500\, \GeV$}}}\label{fig:mssm}
\end{center}
\end{figure}

\begin{table}[htb]
    \begin{center}
        \begin{tabular}{c|cc}
            \hline
            scenario & no-mixing &  Small $\alpha_{eff}$  \\ \hline
           $M_{\Azero}\, (\GeV)$ & 400  & 300 \\
            $M_{SUSY}\, (\GeV)$ & 2000 & 800 \\
            $\mu\, (\GeV)$ & 200 & 2000  \\
            $X_t \equiv A_t - \mu/\tan\beta \,(\GeV)$ & $0$ & $-1100$ \\
            $M_2\, (\GeV)$ & 200 & 500 \\
            {$M_3\, (\GeV)$} & 1600& 500 \\
        \end{tabular}
    \end{center}
    \caption{\footnotesize{MSSM parameter settings corresponding
      to two benchmark scenarios, as defined in Ref.~\cite{benchmarks}. GUT relations {between
      the electroweak gaugino soft SUSY-breaking masses}, as well as universal trilinear couplings ($A_t=A_b=A_\tau$), are assumed.}}
    \label{tab:mssmparam}
\end{table}
{It is precisely this kind of effects that were reported in the
original MSSM calculations for {single Higgs boson} production, cf.
Ref.~\cite{2phot_mssm,previousmssm}. The scenarios considered therein, however,
become problematic} when they are revisited in the light of the
current constraints on the MSSM parameter space. The presence of
light stops, {combined with a rather large trilinear coupling $A_t$,
induces sizable one-loop corrections to the light \CP-even Higgs
boson mass $M_{\hzero}$, which easily clash} with the limits on the
phenomenologically excluded mass regime. By a similar token, light
stops tend to be disfavored from either indirect restrictions
(mainly from $\bsg$ constraints\,\cite{Misiak:2006zs}) and from the
direct searches conducted at the Tevatron, and currently underway at
the LHC \cite{cms}. To be sure, many of the theoretically best
motivated realizations of SUSY tend to accomodate a squark spectrum
with masses heavier than a few hundred GeV -- {this is indeed} the
reason why relatively heavy squarks are ubiquitous in the standard
benchmark points defined in the literature (viz. the Les Houches
~\cite{benchmarks}  or the SPS convention \cite{sps}).

\begin{figure}[t]
\begin{center}
\includegraphics[scale=0.75]{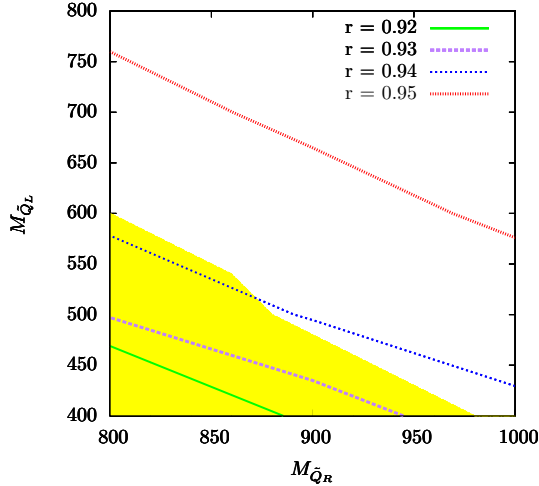}
\caption{\footnotesize{ Ratio
$g_{\hzero\Pphoton\Pphoton}/g_{H\Pphoton\Pphoton}$ in the
$M_{\tilde{Q}L}$ - $M_{\tilde{Q}R}$ plane, cf. Eq.\,
(\ref{eq:sfermionmass2}), in which the separate variation of the
left- and right-handed squark soft SUSY-breaking masses gives rise
to an explicit mixing in the squark chiral sector -- see the text
for details. The remaining MSSM parameters are set along as follows:
$\tan\beta = 2$, $M_{\Azero} = 600 $ GeV, $\mu = 500 $ GeV, $A_t =
1800$ GeV, $M_2 = 500$ GeV. GUT relations {between $M_1$ and $M_2$},
as well as universal trilinear couplings ($A_t=A_b=A_\tau$), are
assumed. The shaded region is ruled out by $\bsg$.}
}\label{scanmssm}
\end{center}
\end{figure}

With these provisos in mind, let us now present the results of our
own (fully updated) calculation of the single MSSM Higgs boson
production at a photon collider, $\Pphoton\Pphoton \to h \, (h =
\hzero, \Hzero, \Azero)$,  by taking into account, in particular,
the current mass bounds stemming from direct SUSY particle searches
at the LEP and Tevatron \cite{pdg}, and most significantly the
presently allowed Higgs boson mass range \cite{higgsbounds}. Further
restrictions, such as the compliance with the limits imposed by
$\bsg$\,\cite{Misiak:2006zs} and $B_d^0-\bar{B}_d^0$
data\,\cite{superiso}, are also duly taken into account. Worthwhile
mentioning is that, in contrast to the general 2HDM case, here we do
not have severe unitarity bounds because the MSSM Higgs boson
self-couplings are purely gauge. Even so, {the predicted
$\gamma\gamma \to \hzero$ rate in the MSSM} is highly subdued by the
remaining constraints and, overall, it appears rather mild, in the
sense of being highly undifferentiated with respect to the SM case,
whereas the signals for $\Hzero$  and $\Azero$ production are
usually much smaller. A panoramic view of the MSSM results is
presented in Figs.~\ref{fig:mssm} and \ref{scanmssm}.

Let us dwell on these Figures {in more detail}. For example, in
Fig.~\ref{fig:mssm} we survey the total MSSM single Higgs  boson
cross section $\langle\sigma_{\gamma\gamma\to \hzero}\rangle$ at
fixed $\sqrt{s} = 500\,GeV$ as a function of $\tan\beta$, for two
standard benchmark points (cf. Table~\ref{tab:mssmparam}), and we
compare it to the SM yield -- identifying $M_{\PHiggs_{SM}}$ with
$M_{\hzero}$. We have computed in Fig.~\ref{fig:mssm} (bottom
panels) the corresponding mass spectrum for the neutral, \CP-even
states with the help of  \textit{FeynHiggs}\,\cite{feynhiggs}. The
obtained cross sections for $\hzero$ lie very close, though slightly
below, the SM expectations -- similarly to the behavior exhibited by
the 2HDM for those scenarios with small 3H self-couplings. This
translates into few thousand event rates  --  few hundred for
$\Hzero$, and even less for $\Azero$. The profile of
$\langle\sigma_{\gamma\gamma\to\hzero}\rangle$ as a function of
$\tan\beta$ is essentially featureless and is mostly correlated to
the change in the Higgs boson mass. We also notice from the bottom
panels of Fig.~\ref{fig:mssm} that the mass splitting between the
\CP-even Higgs bosons can never mimic the 2HDM situation previously
illustrated in Fig.\,\ref{fig:1}, in which these states could be
simultaneously produced with similar cross-sections. Indeed, we see
that in the MSSM case there is a suppression of the heavy \CP-even
Higgs by roughly one order of magnitude because the behavior of the
cross-section can never be enhanced by a moderately heavier Higgs boson mass,
in contrast to the general 2HDM case. We point out that we have
carried out the same analysis for the other benchmark points defined
in Ref.~\cite{benchmarks} and found very similar phenomenological
trends to those that characterize the \emph{no-mixing} scenario, and
so we will not report on these results in this Letter.

%%%%%%%%%%%%%%%%%%%%%%%%%%%%%%%%%%%%%%%%%%%%%%%%%%%%%%%%%%%%%%

{Let us note that we have called} the ``tail of subleading effects''
in the 2HDM case is also the main source of the MSSM corrections and
proceeds essentially through the same Yukawa, and Yukawa-like,
couplings of the Higgs bosons with the quarks {(here also with the
squarks)}, although in this case the angles $\alpha$ and $\beta$ are
of course tied by the {SUSY relations}\,\cite{hunter}. Thus, in
contrast to the 2HDM, the MSSM is unable to furnish a significant
enhancement or suppression of the ratio $r = g_{\Pphoton\Pphoton
h}/g_{\Pphoton\Pphoton H}^{\rm SM}$ (see Fig.\,\ref{scanmssm}, and
compare it with Fig. \ref{fig:compare1_2hdm}), the reason being the
absence of large 3H self-couplings, and hence the lack of a
mechanism capable of prompting the characteristic interference
pattern that we have singled out for the 2HDM. At the same time the
additional, purely SUSY, contributions to $g_{\Pphoton\Pphoton h}$,
namely the squark-mediated loops (whose enhancing capabilities
originate in the Higgs-squark Yukawa couplings) turn out to be not
so competitive, as they are pulled down by inverse powers of the
SUSY-breaking mass scale, {and further limited by the Higgs and
squark mass bounds and the flavor physics restrictions}. {As a
matter of fact, our updated calculation of $r =
g^{MSSM}_{\Pphoton\Pphoton h}/g_{\Pphoton\Pphoton H}^{\rm SM}$
displays departures from $r=1$ which are typically milder than those
reported in former analyses on the
topic~\cite{2phot_mssm,previousmssm}. Upon sweeping the MSSM
parameter space, we confirm that the prominent regions documented in
the old literature do exist theoretically, although they are no
longer allowed in practice when all current phenomenological
constraints are plugged into the analysis. In particular, the
combination of the Higgs boson mass bounds and the $\bsg$
restrictions turns out to cripple considerably the formerly reported
enhancement power encompassed by the MSSM. This is what we aim at
illustrating in Figure~\ref{scanmssm}, where the evolution of the
relative $\gamma\gamma \hzero$ coupling strength $r$ is explored as
a function of the left-to-right squark mass splitting. For a sizable
Higgs-stop trilinear coupling $A_t = 1800 \,\GeV$, we single out
deviations up to $r \sim -5\%$, which are correlated with the
lightest attainable squark masses -- and the maximum mass splitting
between their chiral components.} We conclude that the MSSM can only
induce rather tempered quantum effects as compared to the 2HDM.

\section{Discussion and conclusions}

In this Letter, we have reported on a comparative study of the
production of a single neutral Higgs boson, $h =
\PHiggslightzero,\,\PHiggsheavyzero,\,\PHiggspszero$ both within the
general two-Higgs-doublet model (2HDM) and in the Minimal
Supersymmetric Standard Model (MSSM). Motivated by the robust handle
on new physics that would represent a precise measurement of the
effective $\Pphoton\Pphoton h$ coupling at a photon collider, we
have computed the single Higgs boson production cross sections
$\langle\sigma_{\gamma\gamma\to \hzero}\rangle(s)$ in the
aforementioned models and compared them to the SM results. In our
study we have applied all known current restrictions on the
parameter spaces of both models coming from unitarity,
perturbativity, custodial symmetry and low-energy flavor physics.
The typical values for the production cross-section of the lightest
\CP-even state $\hzero$ at $\sqrt{s}=500$ GeV fall in the ballpark
of $\sigma\sim\mathcal{O}(10^{-2}) \picobarn$ in both the 2HDM and
the MSSM. In contrast, while the heaviest $\CP$-even state
($\Hzero$) can be produced with similar (even higher) rates in the
2HDM, its cross-section is roughly one order of magnitude depressed
in the MSSM. The next relevant issue is to understand how the extra
degrees of freedom and/or the non-standard dynamical features of
either model, the 2HDM or the MSSM,  may leave a significant imprint
of the new physics, and whether they can give rise to distinctive
signatures. The size of the 3H self-couplings plays a decisive role
here. Depending on the strength of the self interactions between the
charged and the neutral CP-even Higgs bosons in the general 2HDM,
one of the following three situations emerges:

\begin{enumerate}
 \item {Large $\lambda_{h\PHiggs^+\PHiggs^-}$ self-coupling.
If sufficiently enhanced (namely
$\lambda_{h\PHiggs^+\PHiggs^-}\gtrsim10^3$ GeV), this coupling would
induce a large contribution from the charged Higgs boson mediated
loops that would overcome the combined (negative) quantum
effects driven by the fermion and the gauge boson loops. This is the
scenario originally exploited in Ref.~\cite{previous}, {in which the
size of the 3H self-couplings was boosted by a large value of
$|\lambda_5|\gtrsim10$}. However, if one adopts a more conservative
assumption for the unitarity bounds\,\cite{unitarity}, this scenario
becomes unfavored.}

 \item{Moderate $\lambda_{h\PHiggs^+\PHiggs^-}$ at the level of
$\mathcal{O}(10^2)$ GeV. These values are amply permitted by the
more {restrictive unitarity bounds~\cite{unitarity}} and yet produce
a substantial (destructive) interference with the gauge boson and
fermion mediated loops, thence pulling the expected single Higgs
boson rates down to values below the SM expectations, although still
perfectly measurable in many cases. Interestingly enough, both the
scenarios 1) and 2) are theoretically very appealing, as they rely
on a genuine dynamical feature of the 2HDM -- namely the
``Yukawa-like'' nature of the Higgs boson self-interactions and
their enhancing potential -- which is unmatched in the MSSM.}

\item{Small $\lambda_{h\PHiggs^+\PHiggs^-}$, {roughly of
$\mathcal{O}(10)$ GeV}, such that the charged Higgs boson mediated
corrections are relegated to a subleading level. In this case, one
is basically left with the SM-like gauge and Yukawa contributions,
with an additional modulation according to how quarks and gauge
bosons couple to $h = \hzero, \Hzero, \Azero$ in the 2HDM. These
non-standard features translate numerically into $r \equiv
g_{\Pphoton\Pphoton h}/g_{\Pphoton\Pphoton h}^{\rm SM}\lesssim 1$ --
hence a rather mild depletion of the single Higgs boson rate with
respect to the SM one. This situation shows a clear overlap with the
typical picture that we have obtained for the MSSM, where one has,
in addition, the Yukawa-like effects from the Higgs boson
interactions with squarks. In such circumstance there is still a
chance to discriminate the $\gamma\gamma \to h$ signatures triggered
by both models (2HDM and MSSM), most significantly through a
possible correlation of the $\gamma\gamma \to \hzero$ and
$\gamma\gamma \to \Hzero$ processes. Indeed, as SUSY enforces a
relatively large mass splitting between $\hzero$ and $\Hzero$
(cf. bottom panels of Fig.\,\ref{fig:mssm}), it would be
unable to account for {a scenario such as the one addressed} in
Fig.\,\ref{fig:1}, in which the {two \CP-even Higgs bosons} are
produced at similar sizeable rates. Such situation would manifest
through the detection of two back-to-back $b$-jets pointing to two
different scalar resonances whose mass separation could possibly be
resolved by the attainable sensitivity in the dijet invariant mass
reconstruction \footnote{A rough estimate of this sensitivity
provides $\Delta M \sim 2\,\GeV$~\cite{previousmssm}, although a
much better mass resolution should be in principle reachable at a
photon collider, cf. e.g. \cite{deroeck}.}. A signature of this sort
would undoubtedly provide a very strong hint of (non-SUSY) Higgs
physics beyond the SM. In practice, of course, this statement holds
only if we assume a situation similar to that of
Fig.~\ref{fig:overtb2hdm}, in which we spotlight regions where both
$\hzero$ and $\Hzero$ are produced at a rate of order
$1-10\,\femtobarn$, namely regions where neither the
$\hzero\PHiggs^+\PHiggs^-$ nor the $\Hzero\PHiggs^+\PHiggs^-$
self-interactions are large enough to sharpen the destructive
interference with the gauge and fermion-mediated loop corrections.}

\end{enumerate}

A few concluding remarks are in order. On the face of the typical
single Higgs boson rates emerging from direct $\gamma\gamma\to h$
scattering, which lie in the ballpark of a few thousand events per
$500$ $\invfb$ of integrated luminosity (for a center-of-mass energy
range of $\sqrt{s} = 500 - 1000\,\GeV$), it is pretty obvious that
the prospects for Higgs boson detection in a $\gamma\gamma$-collider
are quite encouraging. To start with, let us stress that the single
Higgs-boson final state is to be produced essentially at rest.
Therefore, for $M_h<2M_{W}\lesssim 160\,GeV$, the
corresponding signatures should mostly be in the form of
back-to-back, highly energetic, quark jets ($b\bar{b}$, $c\bar{c}$).
For $M_h>2M_{W}$ and specially for
$M_h>2M_{Z}\simeq 180\,GeV$, instead, signatures with two or
four charged leptons in the final state (from $W^{\pm}\to
\ell^{\pm}+\text{missing energy}$ and, particularly, from $Z\to
\ell^+\ell^-$) should be crystal-clear. To these signatures we
should add the radiative decay $h\to\gamma\gamma$, which will be at
work with the same dynamics as the production $\gamma\gamma\to h$
mechanism. Although its branching ratio is generally small
($\lesssim 10^{-3}$), it could be enhanced significantly in the 2HDM
case\,\cite{Posch10}. With enough statistics on these various
signatures and from the analysis of the invariant mass distribution
of the resulting dijet and dilepton-track signatures, the
measurement of the Higgs boson mass(es) should be attainable with
fairly good accuracy, together with a precise determination of the
effective $g_{\gamma\gamma h}$ couplings (typically for $\hzero$,
and most likely also for $\Hzero$ in the 2HDM).

The new results reported here, despite being based on scenarios
markedly different from the ones emphasized in our previous
study\,\cite{previous}, keep on spotlighting the excellent
opportunities offered by direct $\gamma\gamma$ collisions at future
linac facilities, in particular in the domain of high precision
Higgs boson experiments. After having produced one or more Higgs
bosons, an accurate determination of the effective coupling(s)
$g_{\gamma\gamma h}$ might not only carry undisputed evidence of a
non-standard Higgs boson dynamics, but also a distinctive signature
of its fundamental supersymmetric or non-supersymmetric origin. In
the MSSM case, since  $r=g_{\Pphoton\Pphoton
\hzero}/g_{\Pphoton\Pphoton H}^{\rm SM}$ is expected to be rather
close to $1$ it would be necessary to measure the presence of
additional Higgs states. Fortunately, the SUSY $\gamma\gamma\to h$
yield, even if it became now much more subdued (comparatively to
previous studies\,\cite{2phot_mssm, previousmssm}), is still
sizeable. The main mode is the light $\CP$-even state $\hzero$,
which can be produced with cross-sections that amount to a few
thousand events per $500 \invfb$ of integrated luminosity, whereas
the heavy \CP-even state $\Hzero$ (and in some cases the \CP-odd
one, $\Azero$, as well) can still render a few hundred events. This
shows that a photon-photon collider could help decisively in
escaping the ``cul de sac'' kind of situation in which MSSM Higgs
boson physics might end up at the LHC if the physical parameter
space lies in the infamous (so-called) ``LHC
wedge''\,\cite{haberrev}, namely that region characterized by
$M_{\Azero}> 200$ GeV and intermediate values of $\tan\beta$. Should
Higgs boson events potentially detected at the LHC fall in this
``trap'' of the MSSM parameter space, one could not obviously decide
about the nature of the produced single Higgs boson, as the light
supersymmetric \CP-even state $\hzero$ would then be nearly
indistinguishable from the SM Higgs boson (and at the same time the
heavy Higgs bosons would be virtually undetectable at the LHC
there). Remarkably enough, the MSSM benchmark points we have used
(cf. Table 2 and Fig.~\ref{fig:mssm}) are just in the LHC wedge
region, showing that even in this unfavorable circumstance for the
LHC at least two supersymmetric Higgs bosons could still be
accessible to $\gamma\gamma$ physics in the ILC. Clearly, the unique
opportunity offered by a photon-photon collider for a simultaneous
measurement of additional Higgs bosons, with smaller or similar
rates to the $h^0$ one, would suggest new physics of SUSY or
non-SUSY nature respectively.

%%%%%%%%%%%%%%%%%%%%%%%%%%%%%%%%%%%%%%%%%%%%%%%%%%%%%%%%%%%%%%%
\vspace{0.25cm}
 \noindent
\textbf{Acknowledgments}\,\, The work of JS has been supported in
part by DIUE/CUR Generalitat de Catalunya under project 2009SGR502;
by MEC and FEDER under project FPA2010-20807. This work was also
partially supported by the Spanish Consolider-Ingenio 2010 program
CPAN CSD2007-00042.

%%%%%%%%%%%%%%%%%%%%%%%%%%%%%%%%%%%%%%%%%%%%%%%%%%%%%%%%%%%%%%%%%%%%%%%%
%\newcommand{\JHEP}[3]{{\sl J. of High Energy Physics } {JHEP} {#1} (#2)  {#3}}
\newcommand{\JHEP}[3]{ {JHEP} {#1} (#2)  {#3}}
\newcommand{\NPB}[3]{{\sl Nucl. Phys. } {\bf B#1} (#2)  {#3}}
\newcommand{\NPPS}[3]{{\sl Nucl. Phys. Proc. Supp. } {\bf #1} (#2)  {#3}}
\newcommand{\PRD}[3]{{\sl Phys. Rev. } {\bf D#1} (#2)   {#3}}
\newcommand{\PLB}[3]{{\sl Phys. Lett. } {\bf B#1} (#2)  {#3}}
\newcommand{\EPJ}[3]{{\sl Eur. Phys. J } {\bf C#1} (#2)  {#3}}
\newcommand{\PR}[3]{{\sl Phys. Rept. } {\bf #1} (#2)  {#3}}
\newcommand{\RMP}[3]{{\sl Rev. Mod. Phys. } {\bf #1} (#2)  {#3}}
\newcommand{\IJMP}[3]{{\sl Int. J. of Mod. Phys. } {\bf #1} (#2)  {#3}}
\newcommand{\PRL}[3]{{\sl Phys. Rev. Lett. } {\bf #1} (#2) {#3}}
\newcommand{\ZFP}[3]{{\sl Zeitsch. f. Physik } {\bf C#1} (#2)  {#3}}
\newcommand{\MPLA}[3]{{\sl Mod. Phys. Lett. } {\bf A#1} (#2) {#3}}
\newcommand{\JPG}[3]{{\sl J. Phys.} {\bf G#1} (#2)  {#3}}
\newcommand{\JPCF}[3]{{\sl J. Phys. Conf. Ser.} {\bf G#1} (#2)  {#3}}
\newcommand{\FDP}[3]{{\sl Fortsch. Phys.} {\bf G#1} (#2)  {#3}}
\newcommand{\CPC}[3]{{\sl Com. Phys. Comm.} {\bf G#1} (#2)  {#3}}
%%%%%%%%%%%%%%%%%%%%%%%%%%%%%%%%%%%%%%%%%%%%%%%%%%%%%%%%%%%%%%%%%%%%%%%%


\begin{thebibliography}{99}

\bibitem{pioners} P.W. Higgs, \emph{Phys. Lett.} 12 (1964) 32; \PRL{13}{1964}{508};
F. Englert and R. Brout, \PRL{13}{1964}{321}; G.S Guralnik, C.R.
Hagen and T. W. B. Kibble, \PRL{13}{1964}{585}.

\bibitem{haberrev}{H. Haber, \emph{Present status and Future prospects for a Higgs boson discovery at the
Tevatron and the LHC}, \JPCF{259}{2010}{012017}, {arXiv:1011.1038
[hep-ph]}}.


\bibitem{hunter}J.F. Gunion, H.E. Haber, G.L. Kane and S. Dawson,
\textit{The Higgs hunter's guide}, Addison-Wesley, Menlo-Park, 1990.

\bibitem{Branco11} G. C. Branco \textit{et al.}, \textit{Theory and phenomenology of two-Higgs-doublet models}, arXiv:1106.0034.


\bibitem{HiggsRevs} A. Djouadi, \PR{457 }{2008}{1}; \PR
{459}{2008}{1}; S. Heinemeyer, W. Hollik, G. Weiglein, \PR
{425}{2006}{265}; S. Heinemeyer, \textit{Acta Phys. Polon.} {\bf
B39} (2008) 2673, arXiv:0807.2514 [hep-ph].


\bibitem{ILCPhysics} \textit{ILC Reference Design Report Volume 2: Physics
at the ILC}, {arXiv:0709.1893 [hep-ph]}; \textit{Physics interplay
of the LHC and the ILC}, (G. Weiglein \textit{et al.}), \PR
{426}{2006}{47}.

\bibitem{gammacolliders} See e.g.  V. I. Telnov, \NPPS
{184}{2008}{271}; \textit{Acta Phys. Pol.} {\bf B 37} (2006) 1049;
A. de Roeck, \NPPS{179-180}{2008}{94-103}; B. Badelek \textit{et
al.}, \IJMP {A 19}{2004}{5097}.

\bibitem{2phot_mssm}  B. Grzadkowski, J.F. Gunion, \PLB{294}{1992}{361};
J. F. Gunion, H.E. Haber, \PRD{48}{1993}{5190}; D.L. Borden, D.A.
Bauer, D.O. Caldwell, \PRD{48}{1993}{4018}; M. M\"uhlleitner, M.
Kr\"amer, M. Spira, P. Zerwas, \PLB{508}{2001}{311};  D. M. Asner,
J. B. Gronberg, J.F. Gunion, \PRD{67}{2003}{035009};
 P. Niezurawski, A.F. Zarnecki and M. Krawczyk,
\emph{Acta Phys. Polon.} \textbf{B} 37 (2006) 1187.

\bibitem{previousmssm} S-h. Zhu, C-s. Li, C-s. Gao, \emph{Chin. Phys. Lett.} 15 (1998) 89; M. Krawczyk,
\textit{Photon photon and electron photon physics or physics at photon collider},
{arXiv:hep-ph/0307314}.

\bibitem{previous} N. Bernal, D. L\'opez-Val and J. Sol\`a,
\PLB{677}{2009}{38}, arXiv:0903.4978 [hep-ph].

\bibitem{doublephoton}
F. Cornet and W. Hollik, \PLB{669}{2008}{58}; E. Asakawa, D. Harada,
S. Kanemura, Y. Okada and K. Tsumura, \PLB{672}{2009}{354}; A.
Arhrib, R. Benbrik, C.-H. Chen, and R. Santos,
\PRD{80}{2009}{015010};
%Higgs boson pair production in new physics models at hadron, lepton, and photon colliders.
E. Asakawa, D. Harada, S. Kanemura, Y. Okada, K. Tsumura, \PRD
{82}{2010}{115002}, arXiv:1009.4670;

\bibitem{Posch10} P. Posch, \PLB {696} {2011} {447}, arXiv:1001.1759
[hep-ph]; D. Phalen, B. Thomas, J. D. Wells, \PRD
{75}{2007}{117702}, hep-ph/0612219; A. Arhrib, W. Hollik, S.
Pe\~naranda, M. Capdequi Peyran\`ere, \PLB {579}{2004}{361};
hep-ph/0307391; I. F. Ginzburg, M. Krawczyk, P. Osland,
\text{Nucl.Instrum.Meth.} {\bf A472} (2001) 149, hep-ph/0101229.


\bibitem{SUSYenhanced} J.A. Coarasa, D. Garcia, J. Guasch, R.A. Jim\'enez,
J. Sol\`a, \EPJ {2} {1998} {373}, arXiv:hep-ph/9607485; \PLB {425}
{1998} {329}, arXiv:hep-ph/9711472;  D. Garcia, W. Hollik, R.A.
Jim\'enez, J. Sol\`a,  \NPB {427}{1994}{53}, arXiv:hep-ph/9402341;
S. B\'ejar, J. Guasch, D. L\'opez-Val, J. Sol\`a, \PLB {668} {2008}
{364}, arXiv:0805.0973 [hep-ph], and references therein.

\bibitem{loop1} D. L\'opez-Val and J. Sol\`a, \PRD{81}{2010}{033003}, {arXiv:0908.2898 [hep-ph]};
D. L\'opez-Val and J. Sol\`a, PoS RADCOR2009, 045\, (2010),
arXiv:1001.0473 [hep-ph].

\bibitem{loop2}  D. L\'opez-Val, J. Sol\`a and N. Bernal,
\PRD{81}{2010}{113005},
{arXiv:1003.4312 [hep-ph]}; J. Sol\`a and D. L\'opez-Val,
\FDP{58}{2010}{660}.

\bibitem{Moretti2010} M. Moretti, F. Piccinini, R. Pittau, J.
Rathsman, \JHEP {1011} {097} {2010}, arXiv:1008.0820;
%Production of Light Higgs Pairs in 2-Higgs Doublet Models via the Higgs-strahlung Process at the LHC<
S. S. Bao, Y. L. Wu, \PRD {81}{2010}{075020}, arXiv:0907.3606 [hep-ph].
%Neutral Higgs production on LHC in the two-Higgs-doublet model with spontaneous CP violation.

\bibitem{2HDMenhanced}
J. A. Coarasa, J. Guasch, J. Sol\`a and W. Hollik, \PLB {442}{1998}{326},
hep-ph/9808278.

\bibitem{newflavor} A. Pich, P. Tuz\'on, \PRD{80}{2009}{091702}, {arXiv:0908.1554};
A. Buras, M. V. Carlucci, S. Gori, G. Isidori,
\JHEP{1010}{2010}{009}, {arXiv:1005.5310 [hep-ph]}; 
M. Aoki, S. Kanemura, K. Tsumura, K. Yagyu, \PRD{80}{2009}{015017},
arXiv:0902.4665 [hep-ph].

\bibitem{giancarlo} G. Ferrera, J. Guasch, D. L\'opez-Val and J. Sol\`a,
\PLB{659}{2008}{297}, arXiv:0707.3162 [hep-ph]; PoS RADCOR2007, 043
(2007), arXiv:0801.2469 [hep-ph].

\bibitem{arhrib} A. Arhrib, R. Benbrik and C.-W. Chiang,
\PRD{77}{2008}{115013}, arXiv:0802.0319 [hep-ph]; AIP
Conf.Proc.1006, 112 (2008).

\bibitem{neil} R. N. Hodgkinson, D. L\'opez-Val and J. Sol\`a,
\PLB{673}{2009}{47}, arXiv:0901.2257 [hep-ph].


\bibitem{constraints_general} A. Wahab El Kaffas, P. Osland, O. M. Greid, \PRD{76}{2007}{095001},
{arXiv:0706.2997};
 H. Fl\"acher, M. Goebel, J. Haller, A. H\"ocker, K. M\"onig, J. Stelzer, \EPJ{60}{2009}{543},
{arXiv:0811.0009}; N. Mahmoudi, O. Stal, \PRD{81}{2010}{035016},
{arXiv:0907.1791  [hep-ph]}.

%\bibitem{MSSM} H.P Nilles, \PR{110}{1984}{1}; H.E. Haber, G.L. Kane,
%\PR{117}{1985}{75}.

\bibitem{superiso} F. Mahmoudi, \CPC{178}{2008}{745}, arXiv:0710.2067 [hep-ph];
\CPC{180}{2009}{1579}, arXiv:0808.3144 [hep-ph];
\textit{http://superiso.in2p3.fr}.

 \bibitem{unitarity} S. Kanemura, T. Kubota and E. Takasugi, \PLB{313}{1993}{155}; A. Akeroyd, A.Arhrib, 
 E.-M. Naimi, \PLB{490}{2000}{119}, 
arXiv:hep-ph/0006035. See also Sect. III of Ref.\cite{loop1}.

\bibitem{2hdmcalc} D. Eriksson, J. Rathsman, O. Stal, \CPC{181}{2010}{189},
{arXiv:0902.0851}; \textit{http://www.isv.uu.se/thep/MC/2HDMC/}.

\bibitem{higgsbounds} P. Bechtle, O. Brein, S. Heinemeyer, G. Weiglein,
K. E. Williams, \CPC{181}{2010}{138}, {arXiv:0811.4169,
arXiv:1102.1898}; \textit{http://www.ippp.dur.ac.uk/HiggsBounds}.

\bibitem{feynarts} T. Hahn, \textit{FeynArts 3.2, FormCalc}
and \textit{LoopTools} user's guides, available from
{http://www.feynarts.de}; T. Hahn, \textit{Comput. Phys. Commun.}
\textbf{168} (2005) 78.

\bibitem{compaz} V. I. Telnov, \emph{Acta Phys. Polon.} \textbf{B} 37 (2006) 633;
A. F. Zarnecki, \emph{Acta Phys. Polon.} \textbf{B}34\,(2003)\,2741.


\bibitem{ellis} J. Ellis, M. K. Gaillard, D. V. Nanopoulos,
\NPB{106}{1976}{292}.

\bibitem{Misiak:2006zs} M. Misiak \textit{et al.}
\PRL {98}{2007}{022002}.


\bibitem{cms} The CMS Collaboration, \PLB{698}{2011}{196}, {arXiv:1101.1628 [hep-ex]}.

\bibitem{benchmarks} M. S. Carena, S. Heinemeyer and C. E. M. Wagner, \EPJ{26}{2003}{601},
{arXiv:hep-ph/0202167}.


\bibitem{sps} B. C. Allanach et al., \EPJ{25}{2002}{113}, {arXiv:hep-ph/0202233}.


\bibitem{pdg} K. Nakamura et al. (Particle Data Group), \JPG{37}{2010}{075021}.



\bibitem{feynhiggs}
S. Heinemeyer, W. Hollik and G. Weiglein, \CPC{124}{2000}{76},
{arXiv:hep-ph/9812320}; S. Heinemeyer, W. Hollik and G. Weiglein,
\EPJ{9}{1999}{343}, {arXiv:hep-ph/9812472}; G. Degrassi, S.
Heinemeyer, W. Hollik and P. Slavich, \EPJ{28}{2003}{133},
{arXiv:hep-ph/0212020}; M. Frank et al., \JHEP{02}{(2007)}{047},
{arXiv:hep-ph/0611326}.

\bibitem{deroeck} A. de Roeck, \NPPS{126}{2004}{386-395}.









\end{thebibliography}
\end{document}